\documentclass[prd,eqsecnum, %showpacs,
nofootinbib,superscriptaddress]{revtex4}

\global\arraycolsep=2pt
\usepackage{stmaryrd}
\usepackage{amsmath}
\usepackage{amssymb}
\usepackage{graphicx}
\usepackage{textcomp}
\usepackage{calrsfs}
\usepackage{yfonts}
\usepackage{bm}
\newcommand{\ben}{\begin{equation*}}
\newcommand{\een}{\end{equation*}}
\newcommand{\bean}{\begin{eqnarray*}}
\newcommand{\eean}{\end{eqnarray*}}

\newcommand{\nn}{\nonumber}
\newcommand{\be}{\begin{equation}}
\newcommand{\ee}{\end{equation}}
\newcommand{\bea}{\begin{eqnarray}}
\newcommand{\eea}{\end{eqnarray}}

\DeclareMathOperator{\Tr}{Tr}

\begin{document}

\title{Casimir Self-Entropy of an Electromagnetic Thin Sheet}

\author{Yang Li}
  \email{liyang@ou.edu}
  \affiliation{H. L. Dodge Department of Physics and Astronomy, University of Oklahoma, Norman, OK 73019 USA}
\author{K. A. Milton}
  \email{kmilton@ou.edu}
  \affiliation{H. L. Dodge Department of Physics and Astronomy, University of Oklahoma, Norman, OK 73019 USA}
\author{Pushpa Kalauni}
  \email{pushpa@ou.edu}
  \affiliation{H. L. Dodge Department of Physics and Astronomy, University of Oklahoma, Norman, OK 73019 USA}
\author{Prachi Parashar}
  \email{prachi@nhn.ou.edu}
  \affiliation{H. L. Dodge Department of Physics and Astronomy, University of Oklahoma, Norman, OK 73019 USA}
  \affiliation{Department of Physics, Southern Illinois University-Carbondale, Carbondale, IL 62091 USA}

\begin{abstract}
Casimir entropies due to quantum fluctuations in the interaction
between electrical bodies can often be negative,
either caused by dissipation or by geometry.  Although generally such entropies
vanish at zero temperature, consistent with the third law of thermodynamics (the Nernst heat theorem),
there is a region in the space of temperature and
separation between the bodies where negative entropy occurs, while positive
interaction entropies arise for large distances or temperatures. Systematic studies on this phenomenon
in the Casimir-Polder interaction between a polarizable
nanoparticle or atom and a conducting plate in the dipole approximation have been given recently. Since
the total entropy should be positive according to the second law of thermodynamics,
we expect that the self-entropy of the bodies would be sufficiently
positive as to overwhelm the negative interaction entropy. This expectation, however, has not been explicitly verified.
Here we compute the self-entropy of an
electromagnetic $\delta$-function plate, which corresponds to a perfectly
conducting sheet in the strong coupling limit. The transverse
electric contribution to the self-entropy is negative, while the transverse magnetic
contribution is larger and positive, so the total self-entropy is positive.
However, this self-entropy vanishes in the strong-coupling limit. In that
case, it is the self-entropy of the nanoparticle that
is just sufficient to result in a nonnegative total entropy.
\end{abstract}
\date\today
\maketitle

\section{Introduction}
\label{sec:intro}
Negative Casimir entropies were first encountered in
modeling the electrical properties of a metal plate including dissipation due
to finite conductivity \cite{most,brevik,njp}.  It was found that although
the Nernst heat theorem is satisfied, in that the entropy vanishes as the
temperature approaches zero, signalling the existence of a single ground state,
there was an intermediate region of separation between two metal plates and of
temperature in which the entropy was negative.  It was argued that this was
of no serious concern, although perhaps surprising, because it is only the
total entropy, which includes the self-entropies of the two bodies and that of the environment, that must be positive.

\begin{figure}
  \includegraphics{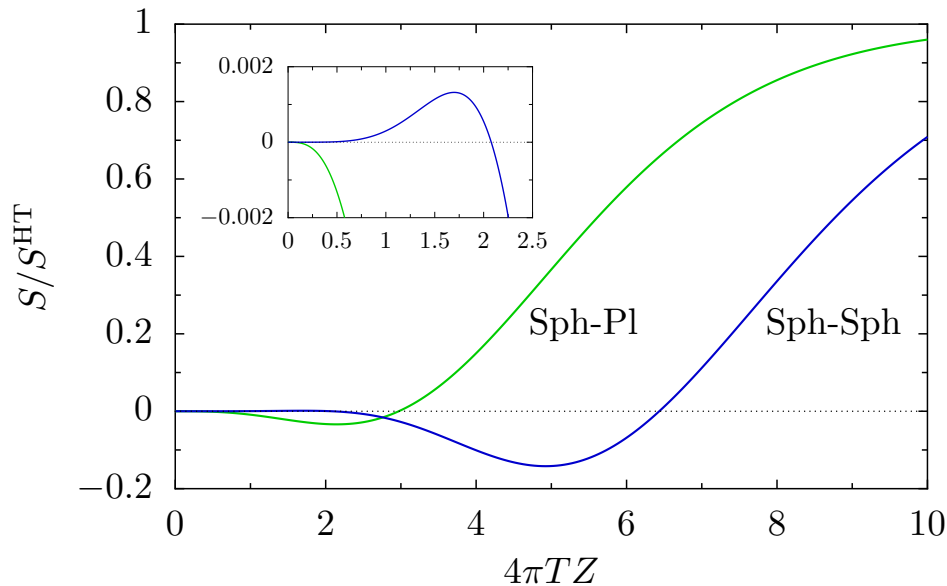}%[width=0.45\textwidth]
  \caption{\label{fig1}
The entropy of interaction between a sphere and a plane (Sph-Pl) and
between two spheres (Sph-Sph) normalized with respect to the corresponding
high-temperature values is displayed as a function of the product of
distance $Z$ and temperature $T$. The entropy has been evaluated
within the dipole and single-scattering approximation \cite{Milton:2016tmr}.
The inset shows the behavior of the
entropy for small $TZ$. We call the negative entropy region perturbative
for the sphere-plane configuration and nonperturbative for the sphere-
sphere configuration.}
\end{figure}

Somewhat later, it was noticed that negative Casimir entropies also emerged
geometrically even for dissipationless materials. For example, negative
entropy occurred in the interaction between polarizable atoms
and conducting plates \cite{most1}. The same phenomenon was observed
between a perfectly conducting sphere and a perfectly conducting plane \cite{canaguier-prl,canaguier-pra},
and between two perfectly conducting spheres \cite{rodriguez-prb,rodriguez-qfext}.
The negative entropy phenomenon was dominated by the dipole approximation, already in the
single-scattering approximation, which led to systematic studies on the
effect for the Casimir-Polder interaction between an anisotropic polarizable
atom or nanoparticle and a conducting plate, or between two such nanoparticles \cite{milton,ingold,umrath}.
The appearance of negative entropy was common, nearly ubiquitous, even between perfect conductors,
as shown in Fig.~\ref{fig1} taken from Ref.~\cite{Milton:2016tmr}.
Here we see the interaction entropy between a sphere and
a plane, and between two identical spheres, calculated in the single-scattering
dipole approximation. When the separation distance $Z$ times the temperature $T$ is of order unity (in natural units),
there is a region where the entropy is negative. For the sphere-sphere case
at room temperature, the entropy is most negative at a readily measurable
distance of a few $\mu$m. The behavior of the perfectly conducting
sphere-plate entropy for small separations is
\be
S_{\rm sp}(T)\sim-\frac4{15}(\pi a T)^3,\quad 4\pi Z T\ll 1,\label{splt}
\ee
where $a$ is the radius of the sphere.

Again, the supposition was that although this negative entropy region is
surprising, it does not violate any fundamental physical principle, because
it is compensated by larger effects, such as the entropy of the vacuum and
that due to the bodies themselves. The purpose of this paper is to
investigate the latter, which has many interesting features in its own right.
The self-energy of a body possesses many well-known divergences, as
does the free energy of a body at finite temperature, but the entropy should be
finite and unambiguous. How this comes about is nontrivial and certainly
the sign cannot be ascertained a priori.

%The organization of this paper is as follows.
%At finite
%temperature the Planck entropy emerges naturally, while a zero value for
%the energy and momentum is achieved only for a regulator symmetric in
%four dimensions.
A model for a thin conducting plate, which we call an electromagnetic
$\delta$-function plate \cite{Parashar:2012it}, is considered in Sec.~\ref{sec3}.
In the strong coupling limit,
this corresponds to a perfectly
conducting surface with zero skin depth.
In Sec.~\ref{sec4}, we investigate the strong and weak coupling limits
of the entropy, which are analyzed in more detail in Sec.~\ref{sec5}
where we calculate the transverse electric (TE) and transverse magnetic (TM) contributions separately.
For this end,
we adopt a plasma model for the dispersive character of the coupling.
Although it is not analytic at the origin, we obtain the TE part of
the free energy by expanding in powers of the coupling (weak coupling expansion). The entropy,
unlike the free energy, is finite and expressed in terms of an
explicit function of the coupling divided by the temperature. In contrast,
for the TM part, the natural expansion is in inverse powers of the coupling (strong coupling expansion). Again,
a finite closed-form expression for the entropy emerges. The results are that
the TE contribution to the entropy (which corresponds
to a scalar $\delta$-function plate---a Dirichlet plate in strong coupling)
is always negative, while the TM contribution is always positive and
larger than the magnitude of the TE part. The total self-entropy
of the plate is thus positive, except in the strong-coupling limit, where the sum
of the two terms vanishes.
The more realistic Drude model for dispersion is then briefly discussed.
Since the entropy of the plates vanishes in strong coupling, we compute, in Sec.~\ref{sec7},
the self-entropies of a polarizable particle. When this is realized as a conducting sphere, the
negative entropy found in the interaction between a perfectly
conducting sphere and a plane is exactly canceled. The general problem of the
entropy of an electromagnetic $\delta$-function sphere will be treated
in a subsequent publication. Some concluding remarks are offered in
Sec.~\ref{sec:concl}.
In this paper, we utilize a point-splitting regulation method, which is illustrated in Appendix~\ref{sec2} by
reconsidering the old problem \cite{christensen} of the vacuum expectation value of the stress
tensor in empty Minkowski space.

We use natural units $\hbar=c=k_B=1$, and Heaviside-Lorentz electromagnetic units except for the polarizabilities which are in Guassian units.

\section{Electromagnetic $\delta$-function plate}
\label{sec3}
We start from the general expression for the free energy with the vacuum energy
subtracted,
\be
F=-\frac{T}2\sum_{m=-\infty}^\infty \Tr\ln\bm{\Gamma}\bm{\Gamma}_0^{-1},
\ee
where the trace is over spatial coordinates and internal variables (tensor
indices). The trace depends on the imaginary Matsubara frequency
$\zeta_m=2\pi m T$.
The Green's dyadic $\bm{\Gamma}$ satisfies
\be
\bigg[-\frac1{\omega^2_m}\bm{\nabla}\times\bm{\nabla}\times-\bm{\varepsilon}(i\zeta_m)\bigg]\bm{\Gamma}=\bm{1}\delta(\mathbf{r}-\mathbf{r}^\prime),
\ee
$\bm{\Gamma}_0$ is the free Green's dyadic and $\bm{\varepsilon}$ is the permittivity tensor of the anisotropic medium. Indentifying the potential $\mathbf{V}=\bm{\varepsilon}-\bm{1}$, the free energy is
\be
F=\frac{T}2\sum_{m=-\infty}^\infty \Tr\ln(\bm{1}-\bm{\Gamma}_0\mathbf{V}).
\label{fe}
\ee

Here, we consider an anisotropic $\delta$-function plate, where
\be
\mathbf{V}=\bm{\lambda}\delta(z)=\mbox{diag}(\lambda_\perp,\lambda_\perp,
\lambda_z)\delta(z).
\ee
In Ref.~\cite{Parashar:2012it}
we showed that $\lambda_z$ must be set equal to zero in
accordance with Maxwell's equations.\footnote{This has been disputed by
Barton \cite{barton} and Bordag \cite{bordag}.  In our case, the value
of the normal component of the coupling does not occur in the expression
for the free energy.}
We, therefore,
write $\lambda=\lambda_\perp$, which could be a function of the imaginary
frequency $\zeta$.
% From that reference, we can also
%read off the
In this approach all we need is the free Green's dyadic, written in this
transverse description,
\be
\bm{\Gamma}_0(\mathbf{r,r'})=\int\frac{(d\mathbf{k}_\perp)}{(2\pi)^2}
e^{i\mathbf{k_\perp\cdot(r-r')_\perp}}\bm{\gamma}_0,\quad \bm{\gamma}_0
=(\mathbf{E+H})g_0,\quad g_0(z,z')=\frac1{2\kappa_m}e^{-\kappa_m(z-z')},
\ee
where $\kappa_m=\sqrt{k^2+\zeta_m^2}$, $g_0$ is the free reduced Green's function, while $\mathbf{E}$ and $\mathbf{H}$
are the polarization dyadic operators for the TE and TM modes, respectively.
%in terms of the transverse electric and transverse magnetic Green's functions,
%where $\mathbf{E}$ and $\mathbf{H}$ are polarization operators given below.
%Here the form of the Green's functions depends whether the two $z$ coordinates
%are on the same or opposite sides of the plate:
%\be
%g^{E,H}(z,z')=\left\{\begin{array}{cc}
%zz'>0:& \frac1{2\kappa}\left(e^{-\kappa|z-z'|}+r^{E,H}e^{-\kappa(|z|+|z'|}
%\right),\\
%zz'<0:&\frac1{2\kappa}t^{E,H}e^{-\kappa(|z|+|z'|)}.
%\end{array}\right.
%\ee
 %and the reflection coefficients are
%\be
%r^E=-\frac{\lambda}{\lambda+2\kappa/\zeta^2},\quad r^H=\frac{\lambda\kappa}
%{2+\lambda\kappa},
%\ee
%while the transmission coefficients are
Because $\mathbf{V}$ is a diagonal matrix in the transverse sector, the
polarization operators are effectively trivial.  In the coordinate system
where $\mathbf{k}_\perp$ lies along the $x$ axis,
\be
\mathbf{V}\mathbf{E}=-\lambda\delta(z)\zeta^2\left(\begin{array}{cc}
0&0\\ 0&1\end{array}\right),\quad
\mathbf{V}\mathbf{H}=\lambda\delta(z)\left(\begin{array}{cc}
1&0\\ 0&0\end{array}\right)\partial_z\partial_{z'},
\ee which are orthogonal.
Expanding the logarithm in the free energy (\ref{fe}), and regulating
the divergent sum and integral by point splitting in imaginary time
and transverse space, as in several recent papers
\cite{Milton:2013xia,Milton:2014psa,Milton:2016sev},
yields
%\be
%F=-\frac{T}2\sum_{m=-\infty}^\infty e^{i\zeta_m\tau}
%\int\frac{(d\mathbf{k}_\perp)}{(2\pi)^2}\left[\ln(1-r^H)+\ln(1+r^E)\right]
%=-\frac{T}2\sum_{m=-\infty}^\infty e^{i\zeta_m\tau}
%\int_0^\infty\frac{(d\mathbf{k}_\perp)}{(2\pi)^2}\ln t^H t^E.
%\ee
%Of course, this is a divergent expression.  Following the procedure in
%we regulate by point splitting the temporal and transverse-spatial directions.
the free energy per unit area
\be F=-\frac{T}{4\pi}\sum_{m=-\infty}^\infty e^{i\zeta_m\tau}\int_0^\infty
dk\,k \,J_0(k\delta)\left[\ln\frac2{2+\lambda\kappa_m}+\ln \frac{2\kappa_m}
{2\kappa_m+\lambda\zeta_m^2}\right],\label{fe1p}
\ee
where the first term is the TM contribution and the second is the TE contribution.
The structures appearing in the logarithms are the TM and TE
transmission coefficients, also expressible in terms of reflection
coefficients,
\be
t^E=\frac1{1+\lambda\zeta_m^2/(2\kappa_m)}=1+r^E
,\quad t^H=\frac1{1+\lambda\kappa_m/2}=1-r^H.
\ee

\section{Asymptotic Coupling Behaviors}
\label{sec4}
\subsection{Strong Coupling Limit}\label{sec4a}
It is illuminating, as we shall see in the next Section,
to examine the contributions of the
TE and TM modes separately.  But the composite structure evidently possesses
significant cancelations.  So we first turn to strong coupling, $\lambda\to
\infty$, which represents a perfect conductor. In that limit the
two logarithms in Eq.~(\ref{fe1p}) combine to give $2\ln(2/\lambda\zeta_m)$,
that is, the $\kappa_m$ dependence has cancelled out.  Then, according to
\be
\int_0^\infty dk\,k\,J_0(k\delta)=0,\quad \delta\neq0,\label{dfn}
\ee
the free energy vanishes, which is not true for the TE and TM
modes individually.

In the strong coupling limit, the TE mode term involves the integral
\be
I(\zeta_m,\delta)=\int_0^\infty dk\,k\,J_0(k\delta)\ln\kappa_m=
\frac{d}{d\alpha}\int_0^\infty
dk\,k\,J_0(k\delta)(k^2+\zeta_m^2)^{\alpha/2}\bigg|_{\alpha=0}
=
-\frac{|\zeta_m|}\delta K_1(|\zeta_m|\delta).\label{aclog}
\ee
Of course, the integral given does not exist. However, the second integral
does exist for $\alpha<-1/2$, so we evaluate it analytically there,
differentiate with respect to $\alpha$, and continue
back to $\alpha=0$. The $m=0$ term is to be understand as a limit as $\zeta_m\rightarrow0$. Alternatively, one gets the same result when integrating by parts,
and omitting the divergent surface
term.  This leads to the expression for the free energy in the strong-coupling
limit
\be
F^{\rm TE}=
\frac{T}{2\pi\delta}\sum_{m=0}^\infty{}'|\zeta_m|K_1(|\zeta_m|\delta),
\ee
where the primed sum means the $m=0$ term is counted
with half weight.

We can evaluate this by using the Euler-Maclaurin sum formula around $m=1$ (we
can't expand around $m=0$ since the summand is singular there). If we
let $g(m)=z_m K_1(z_m)$, where $z_m=\zeta_m\delta$, the sum formula reads
\be
\sum_{m=0}^\infty{}'g(m)=\frac12g(0)+\frac12g(1)+\int_1^\infty dm\, g(m)-
\sum_{q=1}^\infty \frac{B_{2q}}{(2q)!}g^{(2q-1)}(1),\label{em}
\ee
where the $B_n$ are Bernoulli numbers.
%Then we find straightforwardly:
%\begin{subequations}
%\bea
%g(0)&=&1,\\
%g(1)&=&1+\frac14(2\pi T\delta)^2(-1+2\gamma+2\ln\pi T\delta)
%+O((2\pi T\delta)^4),\\
%\int_0^\infty dm\,g(m)&=&\frac1{4T\delta},\\
%\int_0^1dm\, g(m)&=&1+\frac1{36}(2\pi T\delta)^2(-5+6\gamma+6\ln \pi T\delta)
%+O((2\pi T\delta)^5),\\
%g'(1)&=&(2\pi T\delta)^2(\gamma+\ln\pi T\delta)+O((2\pi T\delta)^4),\\
%g^{(2q-1)}(1)&=&(2q-4)!(2\pi T\delta)^2,\quad q>1.
%\eea
%\end{subequations}
The first three terms on the right side of Eq.~(\ref{em}) are obvious and
the remainder sum in the Euler-Maclaurin formula (\ref{em})
is asymptotic, which is evaluated by Borel summation according to Ref.~\cite{ellingsen}
\be
\sum_{q=2}^\infty \frac{B_{2q}}{(2q)!}(2q-4)!=\frac12\left[\frac1{36}-\frac
{\zeta(3)}{4\pi^2}\right].
\ee
When these components are added together, we get the free energy and entropy
\be
F^{\rm TE}=\frac1{8\pi\delta^3}+\frac{\zeta(3)}{4\pi}T^3,\ \ S^{\rm TE}=-\frac{3}{4\pi}\zeta(3)T^2<0.\label{TEFEsc}
\ee
The TM strong-coupling entropy exactly cancels this according to Eq.~\eqref{dfn},
which is explicitly verified in Sec.~\ref{sec:tm}.
\subsection{Weak Coupling Limit}
In the weak coupling limit, $\lambda\to0$, the free energy is approximated as
\be
F=\frac{T}{4\pi}\sum_{m=-\infty}^\infty\int_0^\infty dk\,k\,J_0(k\delta)\frac
\lambda2\left(\kappa_m+\frac{\zeta_m^2}{\kappa_m}\right),
\ee
where we only keep the $O(\lambda)$ terms of the combination of
logarithms in Eq.~(\ref{fe1p}).
% is
%\be
%-\ln\left(1+\frac{\lambda\kappa_m}2\right)-\ln\left(1+\frac{\lambda\zeta_m^2}
%{2\kappa_m}\right)\approx -\frac\lambda2\left(\kappa_m+\frac{\zeta_m^2}
%{\kappa_m}\right),
%\ee
%so
%
%The wavenumber integrals are easily carried out:
%\begin{subequations}
%\bea
%\int_0^\infty \frac{dx\,x\,J_0(x)}{\sqrt{x^2+y^2}}&=&e^{-y},\\
%\int_0^\infty dx\,x\,J_0(x)\sqrt{x^2+y^2}&=&-(1+y)e^{-y},\quad y>0.
%\eea
%\end{subequations}
The integrals are easily carried out, leaving us with a single frequency summation:
\be
F=-\frac{T}{4\pi\delta^3}\left(1-\delta\frac{d}{d\delta}-\delta^2
\frac{d^2}{d\delta^2}\right)\sum_{m=0}^\infty{}'\lambda(i\zeta_m)
\cos(\zeta_m\tau)e^{-\zeta_m\delta}.
\ee
Here we recall that the coupling $\lambda$ could well be a function of frequency.
We will at this point assume such dependence is absent (but see below) and
temporarily set $\tau=0$, since the sum converges without
a temporal cutoff. Then we obtain the free energy per unit area $F$ and entropy per unit area $S$
\be
F=-\lambda\frac{\pi^2}{45}T^4,\quad S=-\frac{\partial F}{\partial T}= \lambda\frac{4\pi^2}{45}T^3>0.\label{wcfe}
\ee

According to the above procedure, the TE and TM contributions are
\begin{subequations}
\be
F^{\rm TE}=\lambda\frac{T}{4\pi\delta}\sum_{m=0}^\infty{}'\zeta_m^2
e^{-\zeta_m\delta}=\frac{\lambda}{4\pi^2\delta^4}-\lambda \frac{\pi^2}{60}T^4
+O(\delta^2),
\ee
\be
F^{\rm TM}=-\lambda\frac{T}{4\pi\delta^3}\sum_{m=0}^\infty{}'(1+\delta
\zeta_m)e^{-\zeta_m\delta}=-\frac\lambda{4\pi^2\delta^4}
-\lambda\frac{\pi^2}
{180}T^4+O(\delta^2).
\ee
\end{subequations}
Now each component possesses a divergent free energy, but the entropy in both
cases is positive, and the sum of the two modes yields the finite result
(\ref{wcfe}).

It might seem more sensible to use something like a plasma or Drude model
to describe the frequency dependence of the coupling $\lambda$.  Suppose
$\lambda=\lambda_0/\zeta_m^2$, where $\lambda_0$ is a constant.  Then we
get a different result for the weak-coupling TE contribution,
\be
F^{\rm TE} =\frac{\lambda_0T}{8\pi\delta}\coth\pi T \delta=\frac{\lambda_0}
{8\pi^2\delta^2}+\frac{\lambda_0 T^2}{24}.\label{scalarwc}
\ee
This yields a \emph{negative\/} contribution to the entropy.
There is no weak-coupling expansion for the TM mode (and, as we will
see, not for the TE mode either) in the plasma model---see Sec.~\ref{sec:tm}
below.
\section{Finite Coupling Behaviors}
\label{sec5}
\subsection{Finite coupling--TE mode}
Consider the free energy from the TE mode (which is the same as for a scalar field under the influence
of a $\delta$-function potential) with the plasma model $\lambda=\lambda_0/\zeta_m^2$,
where $\lambda_0$ is constant. The Drude model differs from the plasma model
merely by the omission of the $m=0$ mode from the Matsubara sum.
%We illustrate the surprisingly elaborate procedure
%by calculating first the
%order $\lambda_0^2$ correction to Eq.~(\ref{scalarwc}).
According to Eq.~(\ref{fe1p}) and the plasma model, $F^{\rm TE}$ is
\be
F^{\rm TE}=\frac{T}{2\pi}\sum_{m=0}^\infty{}'\cos(\zeta_m\tau)\int_0^\infty
dk\,k\,J_0(k\delta)\ln\left(1+\frac{\lambda_0}{2\kappa_m}\right).\label{fte}
\ee
We cannot expand in $\lambda_0$, because the expansion is not valid at $m=0$.

The $m=0$ term in Eq.~(\ref{fte}) is
\be
F^{\rm TE}_{m=0}=\frac{T}{4\pi\delta^2}\int_0^\infty dx\,x\,J_0(x)\ln\left(1+
\frac{\lambda_0\delta}{2x}\right).
\ee
When we integrate by parts, the surface term vanishes and we obtain an
answer in terms of modified Bessel functions and Struve functions.  All we
need is the small $\delta$ behavior:
\be
F^{\rm TE}_{m=0}\sim\frac{\lambda_0T}{8\pi\delta}+\frac{\lambda_0^2T}{32\pi}
\left(\ln\frac{\lambda_0\delta}4+\gamma-\frac12\right),\label{ftezero}
\ee
plus corrections which vanish as $\delta\to 0$.

We expand the remainder of Eq.~(\ref{fte}) to second-order in $\lambda_0$ (as an example):
\be
F^{\rm TE}_{m\ne0}\approx\frac{T}{2\pi}\sum_{m=1}^{\infty}
\cos(\zeta_m\tau)\left[
\frac{\lambda_0}{2\delta}\int_0^\infty \frac{dx\,x\,J_0(x)}
{\sqrt{x^2+\zeta_m^2\delta^2}}
-\frac{\lambda_0^2}8\int_0^\infty
\frac{dx\,x\,J_0(x)}{x^2+\zeta_m^2\delta^2}\right].\label{expfe}
\ee
The two integrals here give convergence factors for the remaining sum on $m$:
\be
\int_0^\infty \frac{dx\,x\,J_0(x)}{\sqrt{x^2+\zeta_m^2\delta^2}}=e^{-|\zeta_m|
\delta},\quad\int_0^\infty \frac{dx\,x\,J_0(x)}{x^2+\zeta_m^2\delta^2}=
K_0(|\zeta_m|\delta).\label{ints}
\ee
The order $\lambda_0$ term is immediately evaluated in terms of a hyperbolic
cotangent, expanded to read, when combined with the corresponding $m=0$ term
from Eq.~(\ref{ftezero}),
\be
F^{\rm TE}_{(1)}=\frac{\lambda_0}{8\pi^2}\frac1{\delta^2+\tau^2}
+\frac{\lambda_0 T^2}{24},\label{fte1}
\ee
a slight generalization of Eq.~(\ref{scalarwc}).  Since it seems that
only the spatial cutoff plays an essential role, we will henceforth for
simplicity set $\tau=0$.
The second-order term in Eq.~(\ref{expfe}) is just a sum of Macdonald
functions,
%\be
%F^{\rm TE}_{(2),m\ne0}=-\frac{\lambda_0^2T}{16\pi}\sum_{m=1}^\infty K_0(2\pi
%mT\delta),
%\ee
which is evaluated using the Euler-Maclaurin sum formula (\ref{em}) and then Borel summation as before.
%The
%sum involving Bernoulli numbers and odd-derivatives of the summand is
%asymptotic and evaluated by Borel summation to be
%\be
%\sum_{q=1}^\infty \frac{(2q-2)!}{(2q)!}B_{2q}=1-\frac12\ln2\pi.
%\ee
%Thus the sum over the Bessel functions is for small $\epsilon$
%\be
%\sum_{m=1}^\infty K_0(m\epsilon)=\frac\pi{2\epsilon}+\frac12\left(
%\ln\frac\epsilon2+\gamma\right)-\frac12\ln2\pi,
%\ee
%and then,
Including the $m=0$ term from Eq.~(\ref{ftezero}),
the ``second-order'' term in the free energy is
\be
F^{\rm TE}_{(2)}=-\frac{\lambda_0^2}{64\pi\delta}+\frac{\lambda_0^2T}{32\pi}
\left(\ln\frac{\lambda_0}{2T}-\frac12\right).\label{fte2}
\ee
Evidently the free energy is not analytic near the origin
in either $\lambda_0$ or $T$.

\begin{figure}
\includegraphics{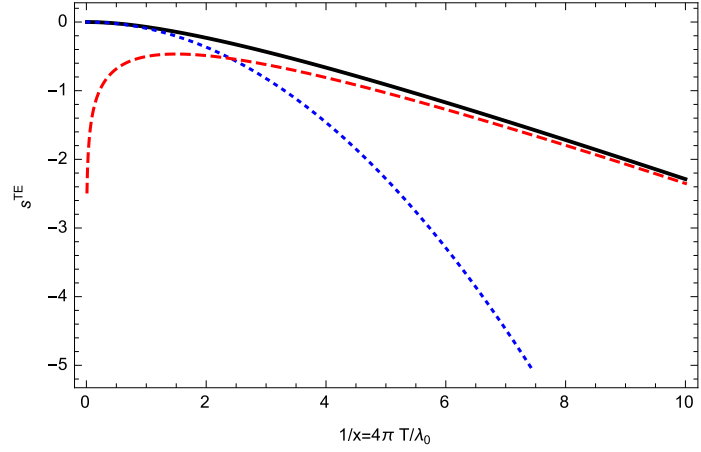}
\caption{\label{figte} The dimensionless
scalar entropy for a single $\delta$-function plate, which is the
same as the TE electromagnetic entropy in the plasma model, shown as a solid
black line, as a function of the temperature divided by the coupling strength.
The dotted blue line shows the
 strong-coupling, low-temperature limit, while the dashed
red line depicts the behavior at high temperature or weak coupling. The entropy
satisfies the third law of thermodynamics, in that the entropy vanishes as the
temperature goes to zero, but it is, surprisingly, always negative.}
\end{figure}

An extension of this method allows us to get a closed form for the scalar
or plasma-model TE free energy. Firstly we integrate Eq.~(\ref{fte}) by parts:
\be
F^{\rm TE}=\frac{\lambda_0 T}{8\pi\delta}\sum_{m=-\infty}^\infty
\int_0^\infty \frac{dx\,x^2\,J_1(x)}{(x^2+\zeta_m^2\delta^2)
(\sqrt{x^2+\zeta_m^2\delta^2}+\lambda\delta/2)}.\label{fteibp}
\ee
The $m=0$ term is just that we found before, in Eq.~(\ref{ftezero}).
We expand the $m\ne0$ terms in (\ref{fteibp}) in powers of $\lambda\delta/2$,
and for the coefficient of $(\lambda\delta/2)^n$ use the integral
\be
\int_0^\infty\frac{dxx^2J_1(x)}{(x^2+y^2)^{(n+3)/2}}=
\frac{|y|^{(1-n)/2}K_{(1-n)/2}(|y|)}{2^{(n+1)/2}\Gamma\left(\frac{n+3}2\right)
},
\quad n>-\frac32.\label{tege}
\ee
The first term ($n=0$)
in this expansion corresponds to the first-order term in
Eq.~(\ref{scalarwc}), while the second term ($n=1$)
gives the second-order term in
Eq.~(\ref{fte2}). The term with $n=2$ gives an explicit logarithm.
 The higher terms are worked out again with the
Euler-Maclaurin formula used for the sum on $m$.
This allows us to evaluate for $n>2$ for small $\epsilon$
\be
\sum_{m=1}^\infty(m\epsilon)^{(1-n)/2}K_{(n-1)/2}(m\epsilon)\sim
\epsilon^{1-n}2^{(n-3)/2}\Gamma\left(\frac{n-1}2\right)\zeta(n-1),
\ee
which gives
\bea
F^{\rm TE}&=&\frac{\lambda_0}{8\pi^2\delta^2}-\frac{\lambda_0^2}{64\pi\delta}
+\frac{\lambda_0 T^2}{24}+\frac{\lambda_0^2T}{32\pi}\left(\ln\frac{\lambda_0}
{2T}-\frac12\right)-\frac{\lambda_0^3}{96\pi^2}\ln2\pi T\delta\nn\\
&&\quad\mbox{}+\frac{\lambda_0 T^2}2\sum_{n=3}^\infty(-1)^n
\left(\frac{\lambda_0}{4\pi T}\right)^n\frac{\zeta(n-1)}{n^2-1}.
\eea
The sum on $n$ here can be carried out in terms of the generalized or
Hurwitz zeta function. % $\zeta(s,a)=\sum\limits_{k=0}^\infty(k+a)^{-s}$.
Although the free energy is divergent, what is of most interest is the finite self-entropy,
\be S^{\rm TE}=-\frac\partial{\partial T}F^{\rm TE}=\frac{\lambda_0^2}{16\pi}
s^{\rm TE}(x),\quad x=\frac{\lambda_0}{4\pi T},
\ee
where we define a dimensionless entropy $s^{\rm TE}(x)$.
The two alternative forms for $s^{\rm TE}$ are
\begin{subequations}
\be
s^{\rm TE}(x)
=
\frac12-\frac1{6x}+\frac{2x}3-\frac{\ln x}{2}
-\frac{3\zeta(3)}{4\pi^2x^2}-\frac{3}{x^2}\zeta^{(1,0)}(-2,1+x)
+\frac{4}x\zeta^{(1,0)}(-1,1+x)+\frac1{x}\zeta^{(1,1)}(-2,1+x)
-2\zeta^{(1,1)}(-1,1+x),
\ee
\be
s^{\rm TE}(x)=
-\frac34-\frac1{3x}+\frac{2x}3+\frac12\ln\frac{x}{2\pi}
-2\ln\Gamma(x)-\frac6{x^2}\psi^{(-3)}(x)+\frac6x \psi^{(-2)}(x),
\ee
\end{subequations}
where $\psi^{(n)}$ are polygamma functions.
It gives the correct limits for large $x$ (strong coupling, low temperature)
\begin{subequations}
\be
x\gg1:\quad s^{\rm TE}(x)\sim -\frac{3\zeta(3)}{4\pi^2x^2}
+\frac1{45x^3}-\frac1{315 x^5}+\frac1{525x^7}-\cdots,\label{largex}
\ee
and for small $x$ (weak coupling, high temperature)
\be
x\ll1:\quad s^{\rm TE}(x)\sim -\frac1{3x}+\frac3{4}-\frac12\ln2\pi x
+\frac{2x}3-\frac{\pi^2}{24}x^2+\cdots.
\label{smallx}
\ee
\end{subequations}
The limit in (\ref{largex}) corresponds to the strong-coupling result
(\ref{TEFEsc}), while the limit in (\ref{smallx})
corresponds to the entropy derived from Eqs.~(\ref{fte1}) and (\ref{fte2}).
Figure \ref{figte} shows how the asymptotic limits are approached by the
exact entropy.  Note that the Nernst heat theorem is satified, in that
the entropy vanishes at zero temperature, but this scalar entropy is always
negative.

\subsection{Finite Coupling--TM mode}
\label{sec:tm}
In the TM case, instead of a weak-coupling expansion, we are naturally led to a
strong-coupling one, again in the plasma model.  It is convenient to subtract
off the leading logarithm term from the first term in Eq.~(\ref{fe1p}), i.e. $F^{\rm TM}=F_1+F_2$,
where, using Eq.~(\ref{dfn}),
\begin{subequations}
\be
F_1=\frac{T}{2\pi}\sum_{m=0}^\infty{}'\cos(\zeta_m\tau)\int_0^\infty dk\,k
\,J_0(k\delta)\ln\left(1+\frac{2\zeta_m^2}{\lambda_0\kappa_m}\right),
\ee
\be
F_2=\frac{T}{2\pi}\sum_{m=0}^\infty{}'\cos(\zeta_m\tau)\int_0^\infty dk\,k
\,J_0(k\delta)\ln\kappa_m.
\ee
\end{subequations}
For simplicity we set $\tau=0$ again.
The integration in $F_2$ is evaluated using
Eq.~(\ref{aclog}). In fact, this term is the negative of the strong-coupling
limit for the TE mode, Eq.~(\ref{TEFEsc}), so we write down without more ado
\be
F_2=-\frac1{8\pi\delta^3}-\frac{\zeta(3)}{4\pi}T^3.
\ee

$F_1$ is the same as the TE free energy
(\ref{fte}) but with the substitution $\lambda_0/2\to 2\zeta_m^2/\lambda_0$.
Thus, here we naturally have a strong-coupling expansion.  The first-order
term in $1/\lambda_0$ is obtained, using Eq.~(\ref{ints}), to be
\begin{subequations}
\label{n123}
\be
F_1^{(1)}=\frac1{\pi^2\lambda_0\delta^4}-\frac{\pi^2}{15}\frac{T^4}{\lambda_0}.
\ee
The second-order term is slightly more complicated, requiring use of the
Euler-Maclaurin formula (\ref{em}) and Borel summation; the result is
\be
F_1^{(2)}=-\frac{9}{4\pi\lambda_0^2\delta^5}-\frac{12\zeta(5)}{\pi\lambda_0^2}
T^5.
\ee
The third-order term can again be done exactly, since it has only exponentials:
\be
F_1^{(3)}=\frac{80}{\pi^2\lambda_0^3\delta^6}-\frac{t^6}{378\pi^2\lambda_0^3},
\ee
where we have now adopted a convenient abbreviation $t=2\pi T$.
\end{subequations}

\begin{figure}
\includegraphics{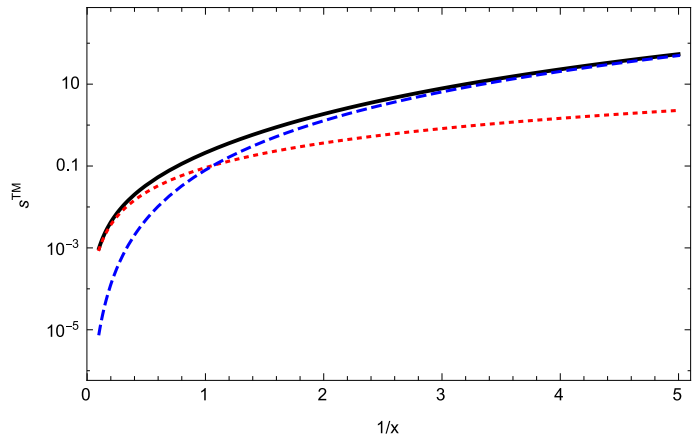}
\caption{\label{tms} Semilog plot of the  transverse magnetic entropy
$s^{\rm TM}$ as a function of $y=1/x=4\pi T/\lambda_0$ for the plasma model.
It is compared with the leading asymptotic expansions for large and small $x$
given in Eqs.~(\ref{tmasy}), in the dotted and dashed lines, respectively.
The TM
entropy is always positive, and overwhelms the negative TE entropy seen in
Fig.~\ref{figte} except for large $x$.}
\end{figure}

In general, we expand $F_1$ in powers of $1/\lambda_0$ as
\be
F_1=\frac{T}{2\pi}\sum_{m=1}^\infty \sum_{n=1}^\infty\frac{(-1)^{n+1}}{n}
\left(\frac{2\zeta_m^2}{\lambda_0}\right)^n\int_0^\infty dk\, k
\frac{J_0(k\delta)}{\kappa_m^n},
\ee
where the integral is evaluated according to
\be
\int_0^\infty\frac{dkkJ_0(k\delta)}{(k^2+\zeta_m^2)^{n/2}}=\delta^{n-2}
\frac{n(|\zeta_m|\delta)^{1-n/2}K_{n/2-1}(|\zeta_m|\delta)}
{2^{n/2}\Gamma(1+n/2)},\quad n>\frac12.
\ee
For a given $n$, we evaluate the $m$ sum using the Euler-Maclaurin sum formula.
In doing so, we encounter the finite sum
\be
-\frac1{n+3}+\frac12+\sum_{k=1}^\infty\frac{B_{2k}}{(2k)!}
\frac{(-n-4+2k)!}{(-n-3)!}
=\zeta(-n-2),\ \ \zeta(1-2k)=-\frac{B_{2k}}{2k}.
\ee
%which vanishes if $n$ is a nonnegative even integer.  For odd integers we have
%\be
%\zeta(1-2k)=-\frac{B_{2k}}{2k}.
%\ee
Thus, the $n$th term in the expansion of the free energy has the form
\be
F_1^{(n)}=\frac{(-1)^{n-1}}{4\pi^2\delta^3}\left(\frac4{\lambda_0\delta}\right)^n
\frac{\Gamma\left(n+\frac12\right)\Gamma\left(\frac{n+3}2\right)}{\Gamma
\left(\frac{n}2+1\right)}-\frac{t^3}{8\pi^2}y^n\frac{B_{n+3}}{n+3}\left(
\frac1{n-2}-\frac1n\right),\quad n>2,
\ee
where $y=1/x=2t/\lambda_0$. Note that the divergent part is independent of
temperature, and hence does not contribute to the entropy. (This expression
agrees with the previous results (\ref{n123}) even for $n=2$,
where an appropriate limit must be understood.)
We calculate the asymptotic sum appearing for the finite part using
Borel summation, namely,
%That is, if
%\be
%f(y)=\sum_{n=3}^\infty \frac{B_{n+3}}{n+3}\frac{y^n}{n-2},\quad
%g(y)=\sum_{n=3}^\infty \frac{B_{n+3}}{n+3}\frac{y^n}{n},
%\ee
%the balance of the free energy is
\be
\sum_{n=3}^\infty F_{1f}^{(n)}
%=-\frac{t^3}{8\pi^2}\sum_{n=3}^\infty\bigg(\frac{B_{n+3}}{n+3}\frac{y^n}{n-2}-\frac{B_{n+3}}{n+3}\frac{y^n}{n}\bigg)
=-\frac{t^3}{8\pi^2}
\int_0^y \frac{du}{u^4}w(u)\left(\frac{y^2}{u^2}-1\right),\label{fng3}
\ee
where
\be
w(u)=\int_0^\infty \frac{dz}z e^{-z}\left[\frac{uz}{e^{uz}-1}-B_0-B_1 u z
-\frac{B_2}2 (uz)^2-\frac{B_4}{4!}(uz)^4\right]=\frac{u}{120}(-60-10u+u^3)
-\ln u-\psi(1/u).
\ee
Here we have regarded the integrand as the analytic continuation of the
Bernoulli series $\sum\limits_{n=0}^\infty [B_{n+6}/(n+6)!] (uz)^n$.  The $u$ integral
in Eq.~(\ref{fng3}) can be carried out, with the explicit result, including
the $F_2$ and the $n=1$ and $n=2$ terms, for the finite part of the TM
free energy in terms of polygamma functions,
\bea
F^{\rm TM}_f&=&-\frac{t^3}{8\pi^2}\bigg[ \frac{3}{2\pi^4x^2}\zeta(5)
+\frac1{2\pi^2}\zeta(3)-\frac{x}{18}-\frac{x^2}{8}-\frac{16x^3}{225}
+\frac{2x^3}{15}\ln x\nn\\
&&\quad\mbox{}+\frac{24}{x^2}\psi^{(-5)}(x)-\frac{24}x\psi^{(-4)}(x)
+10\psi^{(-3)}(x)-2x\psi^{(-2)}(x)\bigg].
\eea
The divergent part of the free energy does not depend on $T$, so does not
contribute to the entropy.  The latter can be written as
\begin{subequations}
\be
S^{\rm TM}=-\frac{\partial F^{\rm TM}}{\partial T}=
\frac{\lambda_0^2}{16\pi}s^{\rm TM},
\ee
where the dimensionless entropy is % (per unit area) is
\be
s^{\rm TM}=\frac{15\zeta(5)}{2\pi^4x^4}+\frac{3\zeta(3)}{2\pi^2x^2}
-\frac1{9x}-\frac18-\frac2{15}x+2\ln\Gamma(x)+\frac{120}{x^4}
\psi^{(-5)}(x)-\frac{120}{x^3}\psi^{(-4)}(x)+\frac{54}{x^2}\psi^{(-3)}(x)
-\frac{14}x\psi^{(-2)}(x).
\ee
\end{subequations}
The behavior of this function for small $x$ (small coupling, high temperature)
is
\begin{subequations}
\label{tmasy}
\be
s^{\rm TM}\sim \frac{15\zeta(5)}{2\pi^4}\frac1{x^4}+\frac{3\zeta(3)}{2\pi^2}
\frac1{x^2}-\frac19\frac1x+\frac18-\frac{2}{15}x+O(x^2),\quad x\ll1,
\ee
while the large $x$ (large coupling, low temperature) expansion is
\be
s^{\rm TM}\sim
\frac{3\zeta(3)}{4\pi^2}\frac1{x^2}+\frac1{15}\frac1{x^3}+\frac{15\zeta(5)}
{4\pi^4}\frac1{x^4}+\frac1{63}\frac1{x^5}+O(x^{-6}),\quad x\gg 1.
\ee
\end{subequations}
The similar appearance of the zeta functions in these two limits is
remarkable.
The entropy is plotted in Fig.~\ref{tms}.
\begin{figure}
\includegraphics{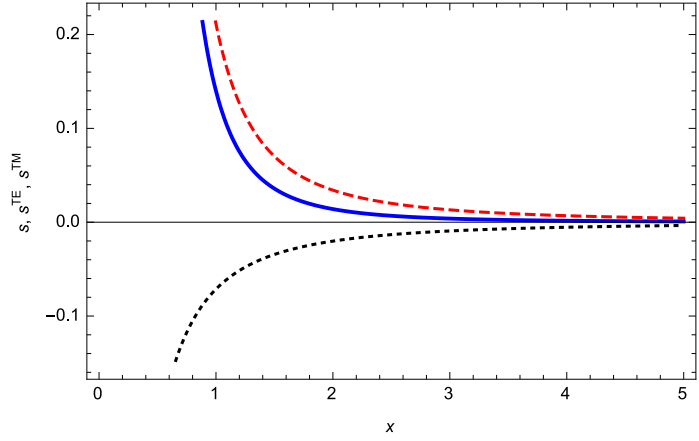}%tm+teentropy-new
\caption{\label{fig:tetm}  The total self-entropy $s=s^{\rm TE}+s^{\rm TM}$ of a electromagnetic
$\delta$-function plate in the plasma model, plotted as a function of
$x=\lambda_0/(4\pi T)$.  Shown for comparison is the TM contribution (red,
dashed line) and the TE contribution (black, dotted line).}
\end{figure}
The comparison between the total self-entropy, $s=s^{\rm TE}+s^{\rm TM}$,
and its TE and TM components is given in Fig.~\ref{fig:tetm}.  It is clear
that the latter dominates, and results in an everywhere positive entropy.
The negative TE entropy contributes equally to the TM entropy
only for large $x$, where the entropy is very small. Therefore, the plasma model
gives a physically satisfactory result: an everywhere positive self-entropy,
which tends to zero at zero temperature.

\subsection{Drude model}
\label{sec6}
The above discussion, consistent with the approach in
Ref.~\cite{Parashar:2012it}, uses a plasma-model type description of dispersion
in the plate.
However, were we to use the Drude model (which for metals is much more
realistic) with the small damping factor $\nu$ and $\lambda=\lambda_0/(\zeta_m^2+\nu\zeta_m)$, the situation is more subtle.
The affect on above
calculations is only substantial for the $m=0$ contributions.
For the TE mode part, it means that the term (\ref{ftezero})
would not be present, which leads,
even in order $\lambda_0$, to a divergent ($1/\delta$) contribution to the
entropy. The TM $m=0$ mode is still vanishing.
The resulting divergent contribution to the entropy appearing in the
Drude model would be
\be
S^D_{\rm div}=\frac{\lambda_0}{8\pi\delta}
+\frac{\lambda_0^2}{32\pi}\ln\delta,
\quad T\gg\nu.
\ee
The Nernst theorem, that the entropy vanishes as $T\to0$, is still
satisfied, but there seems to be a problem for finite temperature, in that
the entropy is no longer finite.
As in the situation with realistic metals, the Drude model, while
undoubtedly more physical, leads to some complications \cite{fqmt11,hbam,bos}.
Investigations along these lines are continuing.

\section{Discussion}
\label{sec7}
As expected, the electromagnetic self-entropy of a thin ($\delta$-function)
plate is positive, although the scalar or TE contribution is negative.
But the total self-entropy of a perfectly conducting plate vanishes, so this
by itself cannot resolve the negative interaction entropy encountered between
a perfectly conducting plate and a perfectly conducting sphere, discussed
in the Introduction and displayed in Fig.~\ref{fig1}. However, we also
must consider the self-entropy of the nanosphere. If we consistently
regard the electric and magnetic polarizabilities, $\alpha$ and $\beta$ respectively, of the sphere as weak,
Eq.~(\ref{fe}) reduces, in the single-scattering approximation, to
\be
F_{b}=-\frac{T}2\sum_{m=-\infty}^\infty \Tr V_{b}
\bm{\Gamma}_0,\quad
V_{b}=4\pi b\delta(\mathbf{r}),\quad b=\alpha,\beta,
\ee
for an isotropic nanosphere at the origin.  We can write the free (vacuum)
Green's dyadic as in Eq.~(\ref{freegd}),
%\be
%\bm{\Gamma}_0(\mathbf{r,r'})=(\bm{\nabla\nabla}-\bm{1}\nabla^2)
%\frac{e^{-|\zeta|R}}{4\pi R}=[\mathbf{\hat R\hat R}v(x)-\bm{1}u(x)]
%\frac{e^{-|\zeta|R}}{4\pi R^3},
%\ee
%with $R=|\mathbf{r=r'}|$, $x=|\zeta|R$, and
%\be
%u(x)=1+x+x^2,\quad v(x)=3+3x+x^2.
%\ee
%Using either of these forms we immediately obtain the electric
which leads immediately to the electric free energy
\be
F_\alpha=\frac{T\alpha}R\sum_{m=-\infty}^\infty \zeta_m^2e^{-|\zeta_m|R}
\bigg|_{R\to0}.
\ee
Here $R$ is the spatial point-splitting quantity. The sum
is written in terms
of a hyperbolic cotangent, which is then expanded for small $R$:
\be
F_\alpha=\frac{2\alpha}{\pi R^4}-\frac{2\alpha}{15}\pi^3T^4,
\ee
and the magnetic free energy has the same expression with $\alpha\to\beta$.
%Once again, the free energy is divergent, but the entropy is finite.
Adding the electric and magnetic free energies for a perfectly conducting
sphere of radius $a$, where $\alpha=-2\beta=a^3$,
we obtain for the entropy of such an object
\be
S_{\rm pcs}=\frac4{15}(\pi a T)^3\label{sesph}.
\ee
 The result
(\ref{sesph}) was first derived by Balian and Duplantier \cite{bd}.
The self-entropy of
a conducting sphere is positive, and precisely cancels the most negative value
of the interaction entropy (\ref{splt}).  So the entropy of the
nanosphere-plate system is always positive, being zero at zero separation.

More generally, if $\alpha\ne\beta$, the sum of the self-entropy of the sphere
and the interacion entropy of the sphere with the plate is
\be
S\ge \frac{16\pi^3}{45}T^3(\alpha+2\beta).
\ee
This will be positive if $\alpha+2\beta>0$, so the perfectly conducting case
is the limiting value to avoid negative entropy.  The Drude model, where $\beta
=0$, would be strictly positive.

For the case of the interaction between two atoms, the situation seems even
more clear-cut.  In that case, for two isotropic atoms separated by a short
distance $Z$ the interaction entropy is given by Ref.~\cite{milton}
\be
S=\frac{(4 \pi T)^5}{5040Z}[11(\alpha_1\alpha_2+\beta_1\beta_2)+
5(\alpha_1\beta_2+\alpha_2\beta_1)].
\ee
Isotropy has resulted in cancellation of the leading term in $(4\pi Z T)$.
Whatever the sign of this term, it always seems much smaller
in magnitude than the sum of
the self-entropies of the nanoparticles.  The size of the interaction
entropy relative to the self-entropy is
\be \frac{T^2}Z \alpha\sim (Ta)^2\frac{a}Z.
\ee
$Ta$ is typically a very small number: for $a=10^{-8}$ cm, at room temperature
$Ta\sim10^{-5}$.  And $a/Z$ is necessarily a small number in order that
the dipole approximation be valid.

\section{Conclusions}
\label{sec:concl}
In this paper, we have calculated the self-entropy of a thin ($\delta$-function)
electromagnetic plate, with a general dispersive coupling $\lambda$.  We
assume a plasma-like dispersion relation and examine the TE and TM mode
in detail, computing the free energy from a weak-coupling expansion in the
first instance and a strong-coupling expansion in the second. In each
cases we get a closed form result for the entropy.
%, expressed in terms of
%polygamma functions.
The TE contribution to the entropy of the plate is
always negative while the TM is positive, and yields a positive total self-entropy.
In strong coupling, the entropy vanishes.

In order to understand how the entropy of the system composed on a polarizable
nanoparticle interacting via quantum fluctuations with a conducting plate,
we must therefore consider the self-entropy of the nanoparticle itself.
At least for the case of a conducting sphere, the later is just sufficient
to render the total entropy positive for all separation distances.

One might have thought that this question is moot, because the entropy
of empty space, discussed in Appendix.~\ref{sec2}, of course overwhelms any small
negative entropy between atoms or between atoms and surfaces.  However, the
latter entropy is quite distinct from the system being studied, so it is
gratifying that positive entropy emerges when the system by itself is
considered.

The observability of negative entropy might also be questioned:  It is not
easy to devise experimental signatures of entropy, with, although physical,
is primarily a theoretical construct.  Perhaps what is more relevant is the
slope of the entropy, or the specific heat,
\be
C_v=T\frac{\partial S}{\partial T}.
\ee
So the signature of something unusual is nonmonotonicity of the entropy, which
of course is exhibited in the interaction between a nanosphere and a conducting
plate.  For further discussion of negative specific heats in this and related
contexts, see Refs.~\cite{spreng,ingold09}.

Elsewhere we will investigate the self-entropy of a $\delta$-function sphere to
complete this self-entropy project. We expect congruence with the
results found here in the strong-coupling (perfectly conducting) limit.
We also would like to explore further the connection between negative
entropy and Casimir repulsion \cite{rep},
both of which involve nonmonotonicity of the free energy.
We hope experimental evidence for both of these phenomena may be soon
forthcoming.

\appendix

\section{Vacuum Stress Tensor}
\label{sec2}
\subsection{Lorentz Invariant Regularization}

The scalar vacuum Green's function, for Euclidean time, is
\be
G_0(R,t_E-t_E')=\int_{-\infty}^\infty \frac{d\zeta}{2\pi}
\frac{e^{i\zeta (t-t')_E-|\zeta|R}}{4\pi R}=\frac1{4\pi^2}\frac1{R^2+(t-t')_E^2},
\ee
where $R^2=(\mathbf{r-r'})^2,\ \ t_E-t_E'=-i(t-t')$ and $\zeta$ is the imaginary frequency, $\omega=i\zeta$.
$G_0$ is just the Euclidean 4D Coulomb propagator,
\be
G_0(\mathcal{R})=\frac1{4\pi^2\mathcal{R}^2},\quad \mathcal{R}^2=R^2+(t-t')_E^2
=R^2-(t-t')^2.
\ee
The stress tensor may be taken to be the canonical one, since the conformal
term will not contribute\footnote{Note, we are working in Minkowski space;
the time regulator is taken to be Euclidean, however.}:
\be
\langle T^{\mu\nu}\rangle=\left[\partial^\mu\partial^{\prime\nu}
-\frac12 g^{\mu\nu}\partial_\lambda\partial^{\prime\lambda}\right]\frac1{4\pi^2
\mathcal{R}^2}.\label{tmunu}
\ee
After differentiation, we get a traceless result coinciding with that of Christensen's
\cite{christensen}
\be
\langle T^{\mu\nu}\rangle=\frac1{2\pi^2 \delta^4}\left[g^{\mu\nu}-4
\frac{\delta^\mu\delta^\nu}{\delta^2}\right],
\quad \delta^\mu=(i\tau,\bm{\rho}),\quad \delta^2=\delta^\mu\delta_\mu=
\rho^2+\tau^2,
\label{vst}
\ee
by taking the coincidence limit $\mathbf{r'}\to \mathbf{r}+\bm{\rho},\ t'\to t+i\tau$ with splittings in time and space $\bm{\rho},
\tau\to0$.
Note that the energy density possesses the familiar time-splitting and
space-splitting Weyl divergent form:
\be
\langle T^{00}\rangle =-\frac1{2\pi^2}\frac{\rho^2-3\tau^2}{(\rho^2+\tau^2)^3}.
\label{weyl}
\ee
In general, the stress tensor (\ref{vst}) is neither diagonal nor
rotationally invariant, which seem unacceptable. Schwinger \cite{ftc} would have
argued that in point splitting, you should average over all directions, so
that
\be
\langle \tau\rho_x\rangle=\langle \rho_x\rho_y\rangle=0,\,\,\mbox{etc.},\quad
\mbox{and}\quad \langle \rho^2_x\rangle=\langle\rho^2_y\rangle=\langle
\rho^2_z\rangle=\frac13\rho^2,\label{3d}
\ee
and then the stress tensor becomes diagonal and has the form characteristic
of radiation,
\be
\langle T^{\mu\nu}\rangle =\frac1{2\pi^2}\frac{\tau^2-\rho^2/3}
{(\tau^2+\rho^2)^3}(g^{\mu\nu}+4\eta^\mu\eta^\nu), \quad \eta^{\mu}=(1,0,0,0).
\ee
This form, for $\tau$ splitting, is
 also given in Christensen \cite{christensen}.
This is still not Lorentz invariant, but it would be if the cutoff is
made 4D rotationally invariant, $\tau^2=\rho^2/3$, just like the spatial point splittings (\ref{3d}).
It is in this sense that the
regulated (consistent with required Lorentz symmetry) vacuum
expectation value of the stress tensor is zero, $\langle T^{\mu\nu}\rangle=0$.
We can turn this argument around and ``understand'' why there is the factor
of $-3$ between the time-splitting and space-splitting results for the Weyl
divergent volume term (\ref{weyl}).

\subsection{Finite Temperature}
For $T>0$ the Green's function is expressed as a sum over Matsubara
frequencies $\zeta_m=2\pi m T$,
\be
G_T(R,t-t')
=T\sum_{m=-\infty}^\infty\frac{e^{i\zeta_m (t-t')_E-|\zeta_m|R}}{4\pi R}
=\frac{T}{8\pi R}\bigg\{\coth \pi T[R-i(t-t')_E]+\coth\pi T[R+i(t-t')_E]\bigg\},
\ee
which reduces to the zero-temperature Green's function $G_0$
as $T\to0$. We compute the energy density using Eq.~(\ref{tmunu}),
\be
\langle T^{00}\rangle_T=-\frac{\pi T^3}{4R}\frac{\cosh\pi T(\rho+i\tau)}{
\sinh^3\pi T(\rho+i\tau)}+(\tau\to-\tau)
\rightarrow
-\frac1{2\pi^2}\frac{\rho^2-3\tau^2}{(\rho^2+\tau^2)^3}
+\frac{\pi^2 T^4}{30},\label{finitet}
\ee
where we have now taken the coincidence limit $R\to\rho$, $t-t'\to i\tau$ with $\rho,\tau\to0$. %Thus we can now expand
%Eq.~(\ref{finitet}) for small arguments of the hyperbolic functions, with
%the immediate result
%\be
%\langle T^{00}\rangle_T=-\frac1{2\pi^2}\frac{\rho^2-3\tau^2}{(\rho^2+\tau^2)^3}
%+\frac{\pi^2 T^4}{30}.
%\ee
The correction to the zero-temperature divergent term is one-half
the usual Planck black-body radiation density.

From the thermodynamic relation $\frac{\partial u}{\partial T}=T\frac{\partial s}{\partial T}$,
we deduce the entropy density $s$ and the free energy density $f$,
\be
s=-\frac{\partial f}{\partial T}=\frac{4\pi^2 T^3}{90}, \quad f=
\frac1{2\pi^2}\frac{3\tau^2-\rho^2}{(\rho^2+\tau^2)^3}-\frac{\pi^2T^4}{90}.
\ee
Other components of the vacuum expectation value of the stress tensor are easily
deduced from the energy density provided we again use the 3-dimensional
averaging procedure. For example,
\be
\langle T_{11}\rangle_T=\frac12(\partial^0\partial^{\prime0}+\partial_1
\partial_1'-\bm{\nabla}_\perp\cdot\bm{\nabla}'_\perp)G_T=\frac13\langle T^{00}
\rangle_T,
\ee
because the spatial derivatives
average to $-\frac13\bm{\nabla}\cdot\bm{\nabla}'$.  Then we deduce
\be
\langle T^{\mu\nu}\rangle_T=\left[\frac1{2\pi^2}\frac{\tau^2-\rho^2/3}{(\rho^2
+\tau^2)^3}+\frac{\pi^2 T^4}{90}\right](g^{\mu\nu}+4\eta^\mu\eta^\nu).
\label{vevstt}
\ee

\subsection{Electromagnetic Vacuum Stress Tensor}
The corresponding construction for the electromagnetic field proceeds in a
very similar manner.  In that case, the free Green's dyadic is
\be
\bm{\Gamma}_0(\mathbf{r,r'})=(\bm{\nabla\nabla-1}\nabla^2)\frac{e^{-|\zeta| R}}
{4\pi R},\quad \mathbf{R=r-r'},\label{freegd}
\ee
so when the identification $i\langle \mathbf{E(r)E(r')}\rangle=\bm{\Gamma}_0(\mathbf{r,r'})$
is made, the point-split regulated energy density, for example, is
\be
u=\int\frac{d\zeta}{2\pi}e^{i\zeta\tau}\frac12\Tr\left(\bm{\Gamma}_0
-\frac1{\zeta^2}\bm{\nabla}\times \bm{\Gamma}_0\times
\overleftarrow{\bm{\nabla}}'\right).
\ee
The electric and magnetic contributions are equal because
\be
\Tr\bm{\Gamma}_0=-\Tr\frac1{\zeta^2}\bm{\nabla}\times\bm{\Gamma}_0\times
\overleftarrow{\bm{\nabla}}'=-\frac{\zeta^2}{2\pi R}e^{-|\zeta|R}.
\ee
The integral over $\zeta$ is easily worked out, with the result
that
\be
u=-\frac1{\pi^2}\frac{\rho^2-3\tau^2}{(\rho^2+\tau^2)^3},
\ee which is twice the scalar result (\ref{weyl}). Since the stress tensor
is traceless, the other components of the vacuum expectation value of the
stress tensor must be twice the scalar result (\ref{vst}) as well.
The finite temperature form must also be twice that given in Eq.~(\ref{vevstt})
because the differential operator appearing in the scalar energy density
construction is
\be
\partial^0\partial^{0\prime}-\frac12\partial_\lambda\partial^{\lambda\prime}
=\frac12\partial^0\partial^{0\prime}+\frac12\bm{\nabla\cdot\nabla}'\to-\zeta^2,
\ee just the multiplier we see in the electromagnetic case.  Indeed, this is
borne out by explicit calculation.

\acknowledgments
We thank the Julian Schwinger Foundation for partial support of this research.
The beginnings of this research were carried out while KAM was on sabbatical
at Laboratoire Kastler Brossel.  He thanks CNRS and the Simons Foundation
for support. He also thanks the OU College of Arts
and Sciences for travel support.
 We thank Gert-Ludwig Ingold, Taylor Murphy, Jacob Tice, and
Steve Fulling for helpful discussions.

\end{document}